\documentclass[twocolumn]{aastex62}
\usepackage[T1]{fontenc}
\usepackage[utf8]{inputenc}
\usepackage[english]{babel}
\usepackage{amssymb}
\usepackage{amsmath}
\usepackage{textgreek}

\newcommand\koi[1]{KOI #1}
\newcommand\kic[1]{KIC#1}
\newcommand\ktwo[1]{K2 #1}

\newcommand\kepler{\textit{Kepler}}

\newcommand\tsup[1]{\textsuperscript{#1}}

\newcommand\supn[1]{\tsup{\ensuremath{#1}}}
\newcommand\expn[2]{\ensuremath{#1 \times 10^{#2}}}

\newcommand\revision[1]{#1}

\makeatletter
\newcommand\DeclareUnit[2]{%
    \@namedef{#1}{\@ifnextchar[{\csname @with@#1\endcsname}{\csname @without@#1\endcsname}}%
    \@namedef{@with@#1}[##1]{\text{#2\supn{##1}}}%
    \@namedef{@without@#1}{\text{#2}}%
}%
\makeatother

\DeclareUnit{km}{\text{km}}
\DeclareUnit{m}{\text{m}}
\DeclareUnit{cm}{\text{cm}}
\DeclareUnit{mm}{\text{mm}}
\DeclareUnit{um}{\text{\textmu m}}
\DeclareUnit{s}{\text{s}}
\DeclareUnit{hr}{\text{hr}}
\DeclareUnit{kg}{\text{kg}}
\DeclareUnit{g}{\text{g}}
\DeclareUnit{pc}{\text{pc}}
\DeclareUnit{au}{\text{AU}}
\DeclareUnit{Pa}{\text{Pa}}
\DeclareUnit{K}{\text{K}}
\DeclareUnit{erg}{\text{erg}}

\newcommand\Jup{\text{Jup}}
\newcommand\MSun{\ensuremath{M_\Sun}}

\newcommand\MEarth{\ensuremath{M_\Earth}}
\newcommand\REarth{\ensuremath{R_\Earth}}
\newcommand\MJup{\ensuremath{M_\Jup}}

\newcommand\eps{\epsilon}
\newcommand\pomega{\varpi}

\newcommand\Om{\Omega}
\newcommand\diff{\mathop{}\!\mathrm{d}}

\newcommand\Plmempty{\ensuremath{P_\ell^m}}
\newcommand\Plm[1]{\Plmempty\!\left(#1\right)}

\newcommand\Pcmb{\ensuremath{P_\textrm{cmb}}}
\newcommand\Pmax{\ensuremath{P_\textrm{max}}}

\newcommand\Ptrans{\ensuremath{P_\textrm{trans}}}
\newcommand\rhotrans{\ensuremath{\rho_\textrm{trans}}}
\newcommand\Chat{\ensuremath{\hat{C}}}
\newcommand\Omhat{\ensuremath{\hat{\Om}}}
\newcommand\Omsqhat{\ensuremath{\hat{\Omega}^2}}
\newcommand\Hhat{\ensuremath{\hat{H}}}
\newcommand\Hhatmax{\ensuremath{{\hat{H}}_\textrm{max}}}
\newcommand\Hmax{\ensuremath{H_\textrm{max}}}

\newcommand\rhomax{\ensuremath{\rho_\textrm{max}}}
\newcommand\rhohat{\ensuremath{\hat{\rho}}}

\newcommand\Phihat{\ensuremath{\hat{\Phi}}}
\newcommand\Psihat{\ensuremath{\hat{\Psi}}}

\newcommand\ellmax{\ensuremath{\ell_\textrm{max}}}

\newcommand\Mstar{\ensuremath{M_\star}}
\newcommand\Mstarhat{\ensuremath{\hat{M}_\star}}
\newcommand\Mp{\ensuremath{M_p}}
\newcommand\Mphat{\ensuremath{\hat{M}_p}}
\newcommand\xphat{\ensuremath{\hat{x}_p}}
\newcommand\xstarhat{\ensuremath{\hat{x}_\star}}
\newcommand\xcmhat{\ensuremath{\hat{x}_\text{cm}}}
\newcommand\ahat{\ensuremath{\hat{a}}}
\newcommand\rhat{\ensuremath{\hat{r}}}

\newcommand\rA{\ensuremath{\hat{r}_{\!_A}}}
\newcommand\rB{\ensuremath{\hat{r}_{\!_B}}}
\newcommand\RA{\ensuremath{r_{\!_A}}}
\newcommand\RB{\ensuremath{r_{\!_B}}}

\newcommand\reltol{\ensuremath{1 \times 10^{-5}}}

\begin{document}
\title{Tidally-Distorted, Iron-Enhanced Exoplanets Closely Orbiting Their Stars}
\author[0000-0002-3286-3543]{Ellen M. Price}
\affiliation{Center for Astrophysics~$|$~Harvard \& Smithsonian, 60 Garden St., Cambridge, MA 02138}
\author{Leslie A. Rogers}
\affiliation{Department of Astronomy, University of Chicago, 5640 South Ellis Ave., Chicago, IL 60637}

\shorttitle{Tidally-Distorted, Iron-Enhanced Exoplanets}
\shortauthors{Price \& Rogers}

\correspondingauthor{Ellen M. Price}
\email{ellen.price@cfa.harvard.edu}

\begin{abstract}
%The origin of Mercury's enhanced iron content is a matter of ongoing debate. 
%{\bf The characterization of rocky exoplanets promises to provide new, independent insights on this topic by constraining the occurrence rate and physical and orbital properties of iron-enhanced planets orbiting distant stars.} 
The transiting planet candidate \koi{1843.03} ($0.6~\REarth$ radius, 4.245 hour orbital period, $0.46~\MSun$ host star) has the shortest orbital period of any planet yet discovered. Here we show, using the first three-dimensional interior structure simulations of ultra-short-period tidally distorted rocky exoplanets, that \koi{1843.03} may be shaped like an American football, elongated along the planet-star axis with an aspect ratio of up to 1.79. Furthermore, for \koi{1843.03} to have avoided tidal disruption (wherein the planet is pulled apart by the tidal gravity of its host star) on such a close-in orbit, \koi{1843.03} must be as iron-rich as Mercury (about 66\% by mass iron compared to Mercury's 70\% by mass iron, \citealt{HauckEt2013JGR}). Of the ultra-short-period ($P_\mathrm{orb} \lesssim 1$~day) planets with physically-meaningful constraints on their densities characterized to date, just under half (4 out of 9) are iron-enhanced. \revision{As more are discovered, we will better understand the diversity of rocky planet compositions and the variety of processes that lead to planetary iron enhancement.} 
%the variety of formation processes that lead to a diversity of rocky planet compositions. 
%origin of the planet Mercury in our Solar System.
\end{abstract}


\section{Introduction}

The compositions of rocky planets reflect a combination of the compositions of their host star, the condensation sequence that concentrates elements heavier than hydrogen and helium into solids, and processing during planet formation and subsequent evolution. To leading order, the Earth is comprised of an iron-dominated core (32\% by mass) and silicate mantle (68\% by mass), with roughly the same relative elemental abundances as in the solar photosphere \citep{Lineweaver&Robles2005ASPCS}. Most rocky exoplanets with measured masses and radii also follow this trend and are consistent with Earth's composition with some scatter \citep{Dressing+2015}. In contrast, Mercury, at 70\% by mass iron \citep{HauckEt2013JGR}, is significantly enhanced in iron relative to solar abundances.

%\textbf{The planet Mercury is the shortest-period planet in our Solar System (though it does not qualify as an ultra-short period (USP) planet). The formation of Mercury remains relatively mysterious, despite our efforts to understand it. It has long been suspected that Mercury must be more iron-rich than the Earth (see \citealt{Ash+1971Science}, for example), because it is smaller than the Earth but has a similar density; this has remained true even with more accurate estimates for the mass and radius \citep{Margot+Mercury}. Earth-based observations of the moment of inertia of the planet indicate differentiation of the core from the mantle and that the core is at least partially liquid \citep{Margot+2012JGR,Wicht+2017arXiv}. The more recent observations, based on data from the {MESSENGER} mission, indicate that Mercury's core may contain sulfur and silicon \citep{Ebel+Mercury}.}

%\textbf{The relatively unusual composition of Mercury and these new results allow us to critically evaluate formation scenarios for this planet. Individual mechanisms are beyond the scope of this introduction, but \citet{Ebel+Mercury} divide current theories into two groups, ``chaotic'' and ``orderly.'' None of the models they enumerate are able to completely explain Mercury's composition. By studying ultra-short period planets in other solar systems and constraining their compositions, we augment the available data and place Mercury in a broader context, which may allow us to better understand it.}

For planets on very short orbital periods ($\lesssim 1$~day), tides can be used to constrain the planets' bulk densities and compositions. Planets in orbit around a star will experience a tidal force, as the day side of the planet feels a stronger attractive gravitational force than the night side. \revision{Planets, by the IAU definition\footnote{\url{https://www.iau.org/static/resolutions/Resolution_GA26-5-6.pdf}} \citep{ExoplanetDefintion}, are sufficiently massive for their self-gravity to overcome their rigid body forces and to achieve hydrostatic equilibrium shapes. As a result,} a tidal field causes an orbiting planet to become elongated in the direction of the planet-star axis \citep[e.g.,][]{dePaterLissauer2010}. If the tidal forces are too strong (the planet is too close to its star), the planet may be tidally disrupted and pulled apart, thus becoming a ring around the host. The minimum distance at which a fluid planet can avoid tidal disruption is called the Roche limit. For an incompressible fluid, this limiting distance is given by
\begin{equation}
    a \simeq 2.44 R_\star \left(\frac{\rho_\star}{\rho_p}\right)^{1/3}
    \label{eqn:roche}
\end{equation}
\citep{Roche1849}, where $a$ is the orbital semi-major axis, $R_\star$ is the stellar radius, $\rho_\star$ is the stellar density, and $\rho_p$ is the planet density. %In Figure~\ref{fig:roche}, we plot this expression for several $\rho_\star$ values over a range of $\rho_p$.
%
% \begin{figure}
%     \centering
%     \includegraphics[width=\linewidth]{fig-roche-theory}
%     \caption{\textbf{Theoretical Roche limit as given by Equation~\ref{eqn:roche}. The dashed line indicates where this limit would predict a planet at the surface of its host star. Three different stellar densities are shown in color, with the scaled orbital radius given as a function of planet density.}}
%     \label{fig:roche}
% \end{figure} 
Following \citet{Rappaport+2013ApJL}, we can rewrite Equation~\ref{eqn:roche} using Kepler's third law to express $a$ in terms of the orbital period $P_\mathrm{orb}$; then, the expression has no dependence on the stellar density, and we find
\begin{equation}
    P_\mathrm{orb} \simeq 12.6~\hr \left(\frac{\rho_p}{1~\g~\cm[-3]}\right)^{-1/2}.
    \label{eqn:roche-mod}
\end{equation}
\revision{The Roche limit is a familiar concept in the context of the rings and satellites of Saturn as well as interacting binary stars.} The discovery of exoplanets that are very close (orbital period, $P_\mathrm{orb} < 1$~day) to their host stars --- found around $0.5\%$ of Sun-like stars \citep{SanchisOjeda+2014ApJ} --- open the opportunity to apply the Roche limit to Earth-mass-scale planets. 

The transiting exoplanet candidate \koi{1843.03} has the shortest orbital period known to date. For \koi{1843.03} to have avoided tidal disruption on such a close-in orbit, previous estimates suggest that it must have a mean density of at least 7~\g~\cm[-3] \citep{Rappaport+2013ApJL}. This density lower limit, however, relies upon interpolating the Roche limits of single-component polytrope models, wherein the pressure $P$ and density $\rho$ within the planet are related by a power-law $P \propto \rho^\gamma$ with constant $\gamma$. These do not accurately capture the density profiles of differentiated rocky bodies with sizes $> 1000$~km. A more accurate calculation of the Roche limit is needed to constrain the composition of \koi{1843.03}.

\revision{In this work, we develop the first self-consistent three-dimensional models for the interior structures of tidally-distorted rocky planets on ultra-short period (USP) orbits ($P_\mathrm{orb} < 1$~day). We apply these models to refine calculations of the Roche limit for USP rocky planets and to explore the diversity of USP planet compositions. 
%In this work, we self-consistently model the interior structures and three-dimensional shapes of tidally-distorted USP rocky planets. and explore the diversity of the resulting compositions.
} The paper is structured as follows: In Section~\ref{sec:methods} we describe the methods used. We outline the primary results in Section~\ref{sec:results} and discuss in Section~\ref{sec:discussion}.

\section{Methods}
\label{sec:methods}

\subsection{Modeling technique}

Calculation of the Roche limit for generic equations of state (EOS) must rely on a numerical solution. Treating the planet as an extended body necessitates computing three forces: the gravitational force from the star, the gravitational force that the planet exerts on itself, and the centrifugal force in the planet's rotating rest frame. The sum of all these forces influence the shape of the planet, which changes the mass distribution and, by extension, the forces on all points inside the planet. There is no simple, analytic way to capture the circular nature of this problem.

We use a relaxation method developed by \citet{Hachisu1986aApJS,Hachisu1986bApJS} to model the three-dimensional structure of USP rocky planets. Starting from an initial guess for the planet density distribution, the method iteratively adjusts the enthalpy and density distribution until a self-consistent solution is reached, within a tolerance of \reltol. We expand the \citeauthor{Hachisu1986bApJS} method to include the gravitational potential of a point source star at a fixed distance from the planet. We also modify the equation of state (which describes how the density of a material varies with pressure) to more accurately capture the behavior of silicate rocks and iron, which have nonzero density at zero pressure. %We have adopted room temperature (300~K) EOSs. Including thermal expansion, which we do not do in this work, will make the lower limits derived on the iron mass fraction of \koi{1843.03} even more severe.

We model two-layer planets consisting of a silicate mantle (enstatite upper mantle and perovskite lower mantle) surrounding an iron core. We simulated more than $280000$ planet configurations over a grid of central pressures $\Pmax$, core-mantle boundary pressures $\Pcmb$, and scaled star-planet orbital separations $a / R_p$. At each grid point, we begin by simulating a nearly spherical planet, self-consistently solving for the host star mass. We then simulate planets that are successively more distorted (elongated along the star-planet axis). Once material begins to fly off the planet, the Roche limit has been surpassed.

\subsection{Coordinate system}

\begin{figure}
    \centering
    \includegraphics[width=0.9\linewidth]{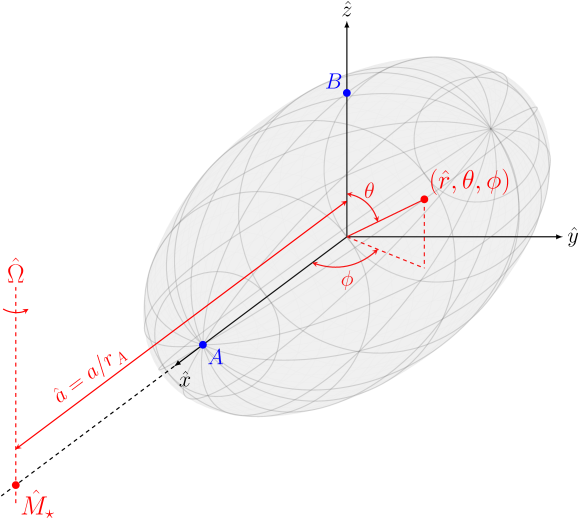}
    \caption{Coordinate system and geometry of the simulation space. The planet center of mass (not necessarily at the origin of the coordinate system) is positioned a distance $\ahat$ along the $x$-axis from a point-like star of mass $\Mstarhat$. The entire system rotates about the stellar axis with angular frequency $\Omega$. Two points, $A$ and $B$, are defined on the surface of the planet such that $A$ lies on the $x$-axis at distance unity from the origin and $B$ lies on the $z$-axis, quantifying the magnitude of the planet's distortion. We use the typical spherical coordinate system with polar angle $\theta$ and azimuthal angle $\phi$, measured from the origin.}
    \label{fig:geometry}
\end{figure}

To model a planet with an unknown shape, we define a three-dimensional coordinate system as shown in Figure~\ref{fig:geometry}, where any point may be specified by a radial coordinate $\hat{r}$, polar coordinate $\theta$ measured from the $z$-axis, and azimuthal coordinate $\phi$ in the $x$-$y$ plane.

Following \citet{Hachisu1986aApJS,Hachisu1986bApJS}, we establish two points, $A$ and $B$, along the $x$- and $y$-axes, respectively, that lie on the surface of the planet. The simulation is conducted in scaled, dimensionless coordinates such that the distance from the origin to $A$ is $\rA \equiv 1$, and, similarly, the scaled distance from the origin to $B$ is $\rB$; we denote the corresponding dimensionful quantities as $\RA$ and $\RB$.

To approximate physical quantities that are continuous over all space, we define a grid of values of $\rhat$, $\mu \equiv \cos{\theta}$, and $\phi$, sampling each quantity at every grid point; for our simulations, we use $N, P, Q = \left( 129, 17, 33 \right)$ divisions in $\rhat$, $\mu$, and $\phi$, respectively. The symmetries inherent in this system allow us to limit the simulation space to $\mu \in \left[ 0, 1 \right]$ and $\phi \in \left[ 0, \pi \right]$; following \citet{Hachisu1986aApJS}, we use $\rhat \in \left[ 0, 16/15 \right]$ to ensure that the planet does not exceed the simulation volume. For a dimensionless physical quantity $\hat{X}$, we use a notation such that $\hat{X}_{i,j,k} = \hat{X}\!\left( \rhat_i, \mu_j, \phi_k \right)$.

The star is treated as a point mass on the $x$-axis with mass $\Mstar$ at a dimensionless distance $\hat{a} = a / \RA$, where $a$ is the radius of the planet's circular orbit, measured from the planet's center of mass to the location of the star. We do \textit{not} make any assumptions about the coordinate of the planet's center of mass, so it does not necessarily coincide with the origin.

The planet rotates about this axis with Keplerian angular velocity given by $\Om^2 = G \left(\Mstar + \Mp\right) / a^3$. Assuming that the planet is tidally locked to the star, we may work in the rest frame of the planet, in which the star is stationary.

\subsubsection{Relaxation method}

A single iteration of the relaxation method begins with a proposal dimensionless density distribution $\rhohat$. We convert the density distribution to a dimensionless enthalpy $\Hhat$ (see Equation~\ref{eqn:dimensionless-enthalpy}). Enthalpy in this context is defined as
\begin{equation}
    H = \int \rho^{-1} \diff P,
    \label{eqn:enthalpy-formal}
\end{equation}
with $\rho$ the density and $P$ the pressure. Enthalpy must meet all of the boundary conditions --- zero enthalpy at $A$ and $B$ and a dimensionless rotation rate $\Omhat$ consistent with Kepler's third law --- according to
\begin{equation}
    \Hhat_{i,j,k} = \Chat - \Phihat_{i,j,k} - \Omsqhat \Psihat_{i,j,k},
\label{eqn:density-to-enthalpy}
\end{equation}
where $\Chat$ is a scalar constant. $\Phihat$ is the total dimensionless gravitational potential, including influence from both the star and the planet, and $\Omsqhat \Psihat$ is the dimensionless centrifugal potential. We then convert the new dimensionless enthalpy to a new dimensionless density distribution. Iterations of this procedure continue until a relative tolerance condition between consecutive iterations is reached. We use a relative tolerance value of $\delta = \reltol$ that must be satisfied for $\Hhat$, $\Chat$, and $\Omsqhat$, such that, between iterations $n$ and $n + 1$,
\begin{equation}
    \left|\max{\left(\Hhat_{n+1} - \Hhat_n\right)} / \max{\left(\Hhat_{n+1}\right)}\right| \leq \delta,
\end{equation}
\begin{equation}
    \left|\left(\Omsqhat_{n+1} - \Omsqhat_n\right) / \Omsqhat_{n+1}\right| \leq \delta,
\end{equation}
and
\begin{equation}
    \left|\left(\Chat_{n+1} - \Chat_n\right) / \Chat_{n+1}\right| \leq \delta.
\end{equation}
These are the same metrics employed by \citet{Hachisu1986aApJS}.

A two-layer planet may be uniquely specified by setting the values of $\Pcmb$, $\Pmax / \Pcmb$, $\ahat$, and $\rB$. For any set of values $\{ \Pcmb, \Pmax / \Pcmb, \ahat \}$, we begin with the largest value of $\rB$ less than 1 and an ansatz dimensionless density distribution $\rhohat$ that is a uniform density ellipsoid with radii $\left( \rA, \rB, \rB \right)$ in the $x$, $y$, and $z$ directions, respectively, ensuring that the ansatz satisfies the boundary conditions. After the system converges, we reduce the value of $\rB$ to the next grid point, effectively increasing the distortion each time, and use the previous converged result as the input ansatz to the next relaxation procedure.

\subsubsection{Potential solver}

By far, the most computationally difficult and expensive component of this method is finding the gravitational potential due to the extended planet itself at every point in space,
\begin{equation}
    \Phi_p\!\left( \mathbf{r} \right) = -G \int\limits_V \frac{\rho\!\left( \mathbf{r^\prime} \right)}{\left| \mathbf{r} - \mathbf{r^\prime} \right|} \diff V^\prime.
\label{eqn:grav-pot-analytic}
\end{equation}
Since the system is symmetric in $y$ and $z$, we may expand and simplify Equation~\ref{eqn:grav-pot-analytic} as

\begin{widetext}
\begin{equation}
    \Phi_p\!\left( r, \mu, \phi \right) = -4 G \sum\limits_{\ell = 0}^{\infty} \sum\limits_{\substack{m = 0 \\ \ell + m~\mathrm{even}}}^{\ell} \!\!\!\epsilon_m \frac{\left( \ell - m \right)!}{\left( \ell + m \right)!} \Plm{\mu} \cos{m \phi} \int\limits_0^\infty \diff r^\prime~f_\ell\!\left( r, r^\prime \right) \int\limits_0^1 \diff\mu^\prime~\Plm{\mu^\prime} \int\limits_0^\pi \diff \phi^\prime~\rho\!\left( r^\prime, \mu^\prime, \phi^\prime \right) \cos{m \phi^\prime}
    \label{eqn:grav-pot-expanded}
\end{equation}
\end{widetext}
where \Plmempty\ are the associated Legendre polynomials,
\begin{equation}
    f_\ell \left(r,r^\prime\right) = 
    \begin{cases}
        {r^\prime}^{\ell+2} / r^{\ell+1}, & \mathrm{if}~r^\prime < r \\
        r^\ell / {r^\prime}^{\ell-1}, & \mathrm{if}~r < r^\prime
    \end{cases},
\label{eqn:f-r-rprime}
\end{equation}
and
\begin{equation}
    \epsilon_m = 
    \begin{cases}
        1, & \mathrm{if}~m = 0 \\
        2, & \mathrm{if}~m \neq 0
    \end{cases}.
\end{equation}
\vspace{1em} % needed to prevent collision
We employ Simpson's rule, following \citet{Hachisu1986bApJS}, in the $\rhat$ dimension but use Gauss-Legendre quadrature in the $\mu$ and $\phi$ dimensions. We develop the discrete, dimensionless equivalent of Equation~\ref{eqn:grav-pot-expanded},
\begin{widetext}
\begin{equation}
    \Phihat_{p;i,j,k} = -4 \sum\limits_{\ell = 0}^{\ellmax} \sum\limits_{\substack{m = 0 \\ \ell + m~\mathrm{even}}}^\ell \epsilon_m \left(\frac{4\pi}{2\ell+1}\right) \widetilde{P}_\ell^m\!\left( \mu_j \right) \cos{m \phi_k}  \times \Lambda_{i,\ell,m}^{\rhat},
\end{equation}
\end{widetext}
where
\begin{equation}
    \Lambda_{i,\ell,m}^{\rhat} = h_{\rhat} \sum\limits_{s = 0}^{N - 1} S_s^N\,f_\ell\!\left( \rhat_i, \rhat_s \right) \Lambda_{s,\ell,m}^\mu,
\end{equation}
\begin{equation}
    \Lambda_{s,\ell,m}^\mu = \sum\limits_{t = 0}^{P - 1}G_t^P\,\widetilde{P}_\ell^m\!\left( \mu_t \right) \Lambda_{s,t,m}^\phi,
\end{equation}
and
\begin{equation}
    \Lambda_{s,t,m}^\phi = \sum\limits_{u = 0}^{Q - 1} G_u^Q\,\rhohat_{s,t,u} \cos{m \phi_u}.
\end{equation}
Here,
\begin{equation}
    \widetilde{P}_\ell^m\!\left(\mu\right) = \sqrt{\frac{2\ell+1}{4\pi} \frac{\left(\ell-m\right)!}{\left(\ell+m\right)!}} P_\ell^m\!\left(\mu\right)
\end{equation}
is the normalized associated Legendre polynomial\footnote{As computed by the function \texttt{gsl\_sf\_legendre\_sphPlm} from the GNU Scientific Library (GSL) \citep{gsl}}. $h_{\rhat}$ is the interval in the $\rhat$ coordinate between successive grid points. The coefficients $S_i^n$ are the alternative composite Simpson's rule coefficients \citep{NumResC}, and the coefficients $G_i^n$ are the fixed-order Gauss-Legendre quadrature weights\footnote{As computed by the function \texttt{gsl\_integration\_glfixed\_point} from the GSL \citep{gsl}}, which depend on the integration interval. Since Gauss-Legendre quadrature is an ``open'' integration scheme, the endpoints of the integration interval in $\mu$ and $\phi$ are excluded; as a matter of computational convenience, we simply inject the endpoints with zero integration weight since they are needed to define the locations $A$ and $B$. We employ the same integration scheme to compute the dimensionless mass of the planet $\Mphat$ and the coordinate of its center of mass $\xphat$. 

Once the planet's center of mass is computed, the coordinate of the star is easily determined by $\xstarhat = \xphat + \ahat$. $\Psihat$ is given by
\begin{equation}
    \Psihat_{i,j,k} = -\frac{1}{2} \pomega_{i,j,k}^2
\end{equation}
where $\pomega_{i,j,k}$ is the distance from the point to the rotation axis; in our case,
\begin{equation}
    \Psihat_{i,j,k} = -\frac{1}{2} \left( \left(\hat{x}_{i,j,k} - \xcmhat \right)^2 + \hat{y}_{i,j,k}^2 \right)
\end{equation}
where
\begin{equation}
    \xcmhat = \frac{\Mphat \xphat + \Mstarhat \xstarhat}{\Mphat + \Mstarhat}
\end{equation}
is the coordinate of the center of mass of the \textit{entire} system. The total gravitational potential $\Phihat$ is given by
\begin{equation}
    \Phihat_{i,j,k} = \Phihat_{p;i,j,k} + \Phihat_{\star;i,j,k}
\end{equation}
where
\begin{equation}
    \Phihat_{\star;i,j,k} = \frac{-\Mstarhat}{\sqrt{\left( \hat{x}_{i,j,k} - \xstarhat \right)^2 + \hat{y}_{i,j,k}^2 + \hat{z}_{i,j,k}^2}}.
\end{equation}
However, at this point, the mass of the star is an unknown. We solve the system of equations that gives the mass consistent with the boundary conditions on the enthalpy and the dimensionless form of Kepler's third law, $\Omsqhat = \left(\Mstarhat + \Mphat\right) / \ahat^3$. Then,
\begin{equation}
    \Omsqhat = -\left( \Phihat|_A - \Phihat|_B \right) / \left( \Psihat|_A - \Psihat|_B \right)
\end{equation}
and
\begin{equation}
    \Chat = \Phihat|_A + \Omsqhat \Psihat|_A.
\end{equation}
Equation~\ref{eqn:density-to-enthalpy} then gives the enthalpy at every point.

We follow \citep{Hachisu1986aApJS} by defining a dimensionless enthalpy $\Hhat$ and its maximum $\Hhatmax$, where
\begin{equation}
    \Hhat \equiv H / G \RA^2 \rhomax.
\label{eqn:dimensionless-enthalpy}
\end{equation}
$\rhomax$ is fixed in the simulation because it can be determined directly from $\Pmax$ and the equation of state. To generate the non-analytic function that maps dimensionless enthalpy to dimensionless density, we finely sample the pressure $P$ from $0$ to $\Pmax$ and calculate the density at each pressure with our equation of state. Equation~\ref{eqn:enthalpy-formal} gives the enthalpy $H$ at each pressure, and $\Hmax$ is just the maximum of these values. Letting $\rhohat \equiv \rho / \rhomax$, we linearly interpolate $\rhohat$ as a function of $H / \Hmax$ and obtain a function which maps scaled, dimensionless enthalpy $\Hhat / \Hhatmax$ to dimensionless density \rhohat.

\subsubsection{Equation of state}

Previous work modelling distorted stars by \citet{Hachisu1986aApJS} assumes a polytropic equation of state, 
\begin{equation}
    \rho\!\left(P\right) = c P^n,
\label{eqn:polytrope-eos}
\end{equation}
where $\rho$ is density and $P$ is pressure. This equation of state is not appropriate for a rocky planet because it does not allow for nonzero density at zero pressure. In this work, we consider only planets with two layers: an iron core and a silicate mantle; our method may be extended to planets with different compositions and any number of layers, however.

\begin{figure}
    \centering
    \includegraphics[width=\linewidth]{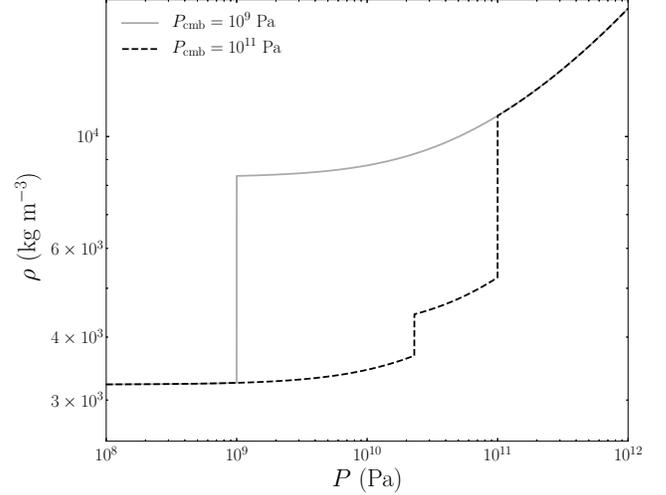}
    \caption{Examples of the piecewise equation of state for two values of the core-mantle boundary pressure, $\Pcmb$. For the lower value of $\Pcmb$, the composition jumps directly from enstatite to iron, whereas, for the higher value, the composition transitions from enstatite to perovskite and then iron.}
    \label{fig:eqnofstate}
\end{figure}

At low pressures ($P \leq \Ptrans = 23\times10^9$~Pa), we apply the Birch-Murnagham equation of state (BME) for enstatite \citep{Seager+2007ApJ}. For $\eta \equiv \rho / \rho_{0,\mathrm{en}}$, we have \citep{Seager+2007ApJ},
\begin{multline}
    P_\mathrm{en}\!\left(\eta\right) = \frac{3}{2} K_{0,\mathrm{en}} \left(\eta^{7/3} - \eta^{5/3}\right) \times \\ \left[1 + \frac{3}{4} \left(K_{0,\mathrm{en}}^\prime - 4\right)\left(\eta^{2/3}-1\right)\right].
    \label{eqn:enstatite-eos}
\end{multline}
Above \Ptrans, we use a tabulated equation of state for perovskite and iron or FeS (depending on the core composition assumed) \citep{Seager+2007ApJ}. The transition between perovskite and the iron-dominated core is defined to occur at a core-mantle boundary pressure \Pcmb. We have adopted room temperature (300~K) EOSs. Including thermal expansion, which we do not do in this work, will make the lower limits derived on the iron mass fraction of \koi{1843.03} even more severe. Figure~\ref{fig:eqnofstate} shows the full equation of state that we have adopted.

The relaxation method (described above) requires the conversion of enthalpy to density, which depends solely on the equation of state. Formally, enthalpy $H$ is given by Equation~\ref{eqn:enthalpy-formal}. The \textit{indefinite} integral that corresponds to substituting Equation~\ref{eqn:enstatite-eos} in Equation~\ref{eqn:enthalpy-formal} is
\begin{widetext}
\begin{equation}
    \widetilde{H}_\mathrm{en}\!\left(\rho\right) = \frac{3 K_0 \rho^{2/3} \left(9 (K_0^\prime - 4) \rho^{4/3} + 7 \left(14 - 3 K_0^\prime\right) \left(\rho \rho_0\right)^{2/3} + 5 (3 K_0^\prime - 16) \rho_0^{4/3}\right)}{16 \rho_0^3}.
\end{equation}
\end{widetext}
The enthalpy as a function of density $\rho$ is then given by
\begin{equation}
    H_\mathrm{en}\!\left(\rho\right) = \widetilde{H}_\mathrm{en}\!\left(\rho\right) - \widetilde{H}_\mathrm{en}\!\left(\rho_{0,\mathrm{en}}\right).
    \label{eqn:enstatite-enthalpy}
\end{equation}

The enthalpy as a function of density for the tabulated equations of state is computed by interpolating a cumulative trapezoidal integration approximating Equation~\ref{eqn:enthalpy-formal}, which we denote $\widetilde{H}_\mathrm{pv}$ and $\widetilde{H}_\mathrm{fe}$ for perovskite and iron, respectively. This gives us, for $P_\mathrm{en}\!\left(\rhotrans\right) = \Ptrans$,
\begin{equation}
    H_\mathrm{pv}\!\left(P\right) = H_\mathrm{en}\!\left(\rhotrans\right) + \left(\widetilde{H}_\mathrm{pv}\!\left(P\right) - \widetilde{H}_\mathrm{pv}\!\left(\Ptrans\right)\right)
    \label{eqn:perovskite-enthalpy}
\end{equation}
and
\begin{equation}
    H_\mathrm{fe}\!\left(P\right) = H_\mathrm{pv}\!\left(\Pcmb\right) + \left(\widetilde{H}_\mathrm{fe}\!\left(P\right) - \widetilde{H}_\mathrm{fe}\!\left(\Pcmb\right)\right).
    \label{eqn:iron-enthalpy}
\end{equation}

\subsection{Model validation}

\revision{To validate our method, we reproduce the classical Roche limit for an incompressible fluid body. For this test case, we achieve $\left|\Delta P_\mathrm{orb}\right| / P_\mathrm{orb} \approx 0.5\%$, where $P_\mathrm{orb}$ is the analytic Roche limiting orbital period (Equation~\ref{eqn:roche-mod}), and $\left|\Delta P_\mathrm{orb}\right|$ is the absolute difference between the analytic expectation and the Roche limit we numerically derive following the method described above. 
%where $\left|\Delta P_\mathrm{orb}\right|$ is the absolute difference between the analytic (Equation~\ref{eqn:roche-mod}) and measured Roche limiting orbital periods and $P_\mathrm{orb}$ is the true Roche limit orbital period.
In our validation experiements, the relative error in $P_\mathrm{orb}$ is observed to be independent of the density of the fluid, as is expected because the simulation is run with a dimensionless, normalized density.}
%We find a virtually constant value for the relative error as a function of density. This can be understood because the simulation is run with a dimensionless, normalized density. Any constant-density body is therefore indistinguishable on the simulation side once scaling has taken place.}

\subsection{Model interpolation procedure}

After all models have been computed, we distill meaningful results by smoothly interpolating within the model grid. When interpolating our model grids, we use the Gaussian process code \texttt{george} \citep{george}. Our chosen kernel is a constant kernel multiplied by a squared exponential kernel. We allow for ``white noise,'' which in this case is not observational but rather computational noise. We also use a convex hull algorithm as a safeguard against extrapolation. This reduces the extent to which our interpolation code can extrapolate outside our models' bounds.

\section{Results}
\label{sec:results}

\subsection{\koi{1843.03}}
\label{sec:KOI1843}

%We apply a new numerical modelling approach, extending a previous method for modelling the structures of rapidly-rotating stars developed by  \citet{Hachisu1986aApJS,Hachisu1986bApJS}, to explore the deformation of differentiated, two-layer rocky planets with iron cores and silicate mantles. 

\begin{figure}
    \centering
    \includegraphics[width=\linewidth]{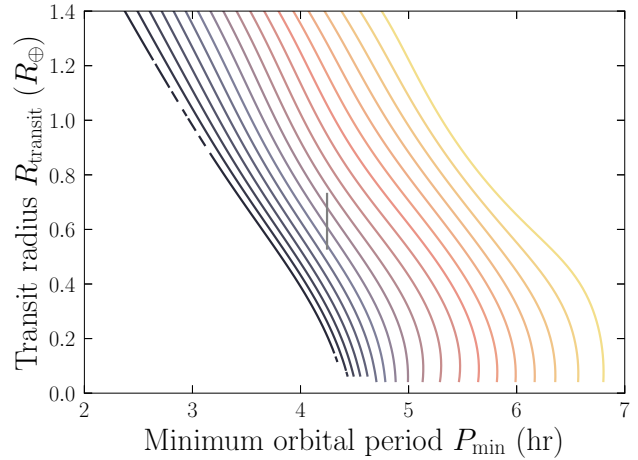}
    \caption{Contours of constant core mass fraction as a function of transit radius and minimum orbital period. The vertical gray line indicates the $1\sigma$ bounds on the transit radius for \koi{1843.03}. Contours are spaced in 5\% intervals in core mass fraction, with darker colors corresponding to high core mass fraction and lighter colors corresponding to low core mass fraction. As expected, denser planets with higher core mass fractions survive to shorter orbital periods.  The limiting orbital period of a pure iron planet is approximately 3.8 hours at $0.5~\MEarth$, 3.6 hours for $1~\MEarth$, and 3.5 hours for $2~\MEarth$. We note that this figure marginalizes over stellar mass, because stellar mass only weakly affects the Roche-limiting minimum orbital period \citep{Rappaport+2013ApJL}. \revision{The stellar density does, however, affect whether the planet can reach its Roche limit before colliding with the star (i.e., $a=R_{\star}$) and thereby the minimum survivable orbital period for the planet.}}
  %  , since, at the orbital period corresponding to the stellar surface, the planet will graze the star and be destroyed; this figure does not take that effect into account.}}
    \label{fig:minperiod-fe}
\end{figure}

\begin{figure}
    \centering
    \includegraphics[width=\linewidth]{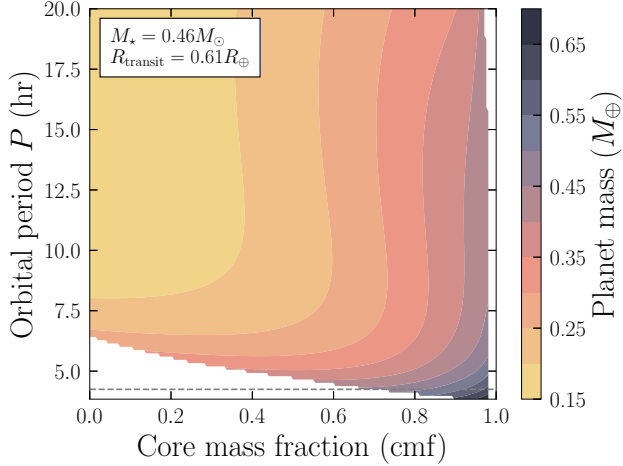}
    \caption{Composition and mass constraints on planets with \koi{1843.03}'s measured transit radius and host star mass. Colored contours show the values of the planet mass, which increases with decreasing orbital period and increasing core mass fraction. The boundary of the colored contours in the lower left-hand corner corresponds to the Roche limit. The dashed gray line indicates the measured orbital period of \koi{1843.03}.}
    \label{fig:contour-koi_1843}
\end{figure}

Our self-consistent 3D models show that \koi{1843.03} must be very iron-rich to avoid tidal disruption on an orbital period of 4.245 hours. Assuming the planet is composed a pure iron core surrounded by a magnesium-silicate mantle, we find that the $R_p = 0.61^{+0.12}_{-0.08}~\REarth$ radius constraints \citep{Rappaport+2013ApJL} imply that \koi{1843.03} must be at least $60_{-8}^{+7}\%$ iron by mass (Figure~\ref{fig:minperiod-fe}). Since rocky planets become compressed to higher densities with increasing size, larger values of the planet radius within the $1\sigma$ range translate into more relaxed lower bounds on the iron mass fraction of the planet. Based on our planet interior models, we expect \koi{1843.03}'s mass to fall in the range $0.32$ -- $1.06~\MEarth$ (Figure~\ref{fig:contour-koi_1843}). %; we show the relevant slices through parameter space in Figure~\ref{fig:slices-mass-koi_1843}. 

%\begin{figure}
%    \centering
%    \includegraphics[width=\linewidth]{fig02-slices-mass-koi_1843}
%    \caption{Mass constraints on \koi{1843.03}, as a function of core mass fraction at three different values for the transit radius; the measured value 0.61~\REarth and the $1\sigma$ limits (0.53~\REarth and 0.73~\REarth). To the left of the plot, we indicate the minimum value of core mass fraction allowed by our models, and, to the right, we indicate the mass bounds.}
%    \label{fig:slices-mass-koi_1843}
%\end{figure}

Figure~\ref{fig:contour-koi_1843} displays interpolated planet masses for a range of orbital periods and core mass fractions in systems consistent with \koi{1843.03}'s host star mass and transit radius. The boundary of the colored contours in the lower left-hand corner corresponds to the Roche limit; as orbital period decreases, the core mass fraction is more tightly constrained. As anticipated, considering a fixed orbital period in Figure~\ref{fig:contour-koi_1843}, increasing the planet's iron mass fraction increases the planet's mass. Less intuitively, at fixed core mass fraction, decreasing the planet's orbital period also leads to an increase in the inferred planet mass. This is due to the tidal distortion of the planet; at shorter orbital periods, the volume of the planet exceeds $4/3\pi R_{\rm transit}^3$ by larger and larger factors. 

\begin{figure}
    \centering
    \includegraphics[width=\linewidth]{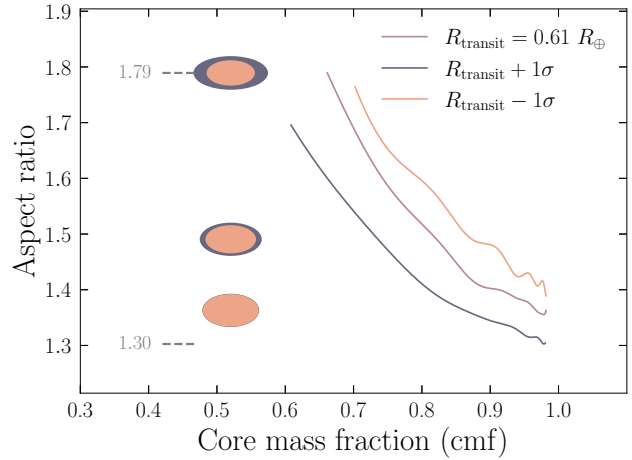}
    \caption{Aspect ratio constraints on \koi{1843.03}, as a function of core mass fraction.  To the left of the plot, we indicate the maximum and minimum values of aspect ratio \revision{supported by the model grid} and show three representative planet cross-sections (assuming ellipsoidal shapes) wherein peach and navy represent the iron-dominated core and silicate mantle, respectively. Results are shown both at the measured transit radius 0.61~\REarth and the $1\sigma$ limits (0.53~\REarth and 0.73~\REarth).}
    \label{fig:slices-aspect-koi_1843}
\end{figure}

As it orbits less than one stellar radius from the host star's photosphere ($a / R_{\star} = 1.9$), \koi{1843.03} will be significantly elongated in the direction of the star due to tidal distortion. Based on the parameters of \koi{1843.03}, our models predict that it must be \textit{at least} $31\%$ longer along the star-planet line than along the perpendicular axes (aspect ratio of about $1.3$), and \revision{our models support a value up to} nearly twice as long along the star-planet line (aspect ratio of almost $1.8$); various possibilities are illustrated in Figure~\ref{fig:slices-aspect-koi_1843}. For comparison, Saturn has an aspect ratio of about $1.1$. 

%The Roche-limiting minimum orbital period and measured transit radius can be used to constrain the core mass fraction of the planet. In Figure~\ref{fig:minperiod-fe}, we show how this may be done for \koi{1843.03}. If the planet were at its Roche limit, then the transit radius puts a reasonable constraint on the core mass fraction. For pure iron cores,

\begin{figure}
    \centering
    \includegraphics[width=\linewidth]{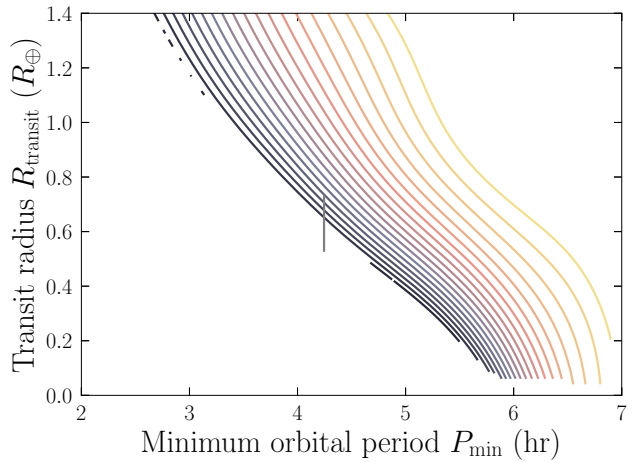}
    \caption{Minimum orbital period as a function of transit radius and composition for planets with cores comprised of pure FeS. The solid gray line indicates the range of possible radii (within $1\sigma$ limits) for \koi{1843.03}, showing that it is, for smaller radii, most likely incompatible with a pure FeS core, as it would be inside the Roche limit. In general, planets with higher density iron cores can survive closer to their host stars than planets with cores polluted by FeS.
    While the limiting orbital period of a pure FeS planet is approximately 4.5 hours at $0.5~\MEarth$, 4.3 hours for $1~\MEarth$, and 4.0 hours for $2~\MEarth$.
    %While the limiting orbital period of a pure iron planet is approximately 3.8 hours at $0.5~\MEarth$, 3.6 hours for $1~\MEarth$, and 3.5 hours for $2~\MEarth$, the corresponding limiting orbital periods for FeS cores are approximately 4.5 hours, 4.3 hours, and 4.0 hours.
    %We note that this figure flattens over the dimension of stellar mass; interpolating to the correct stellar mass, this constraint on \koi{1843.03} becomes even more severe, as it is only compatible with an FeS core if the transit radius is $\gtrsim 2\sigma$ larger than the mean.
    %We note that this figure flattens over the axis of stellar mass, because stellar mass only weakly affects the minimum orbital period \citep{Rappaport+2013ApJL}. 
    This figure is analogous to Figure~\ref{fig:minperiod-fe}, but assumes an FeS core composition instead of Fe.}
    \label{fig:minperiod-fes}
\end{figure}

The Earth's core is not pure iron; it contains an unknown mixture of light elements. To determine what effect these light elements might have, we generated a second grid of models with an EOS appropriate to an FeS core (see Figure~\ref{fig:minperiod-fes}). \koi{1843.03} would need a core mass fraction of at least $80\%$ if it has a core comprised of FeS, but this measurement is only valid for a transit radius of $2\sigma$ greater than the mean. Both pure Fe and pure FeS are end-member core compositions; the true core density of \koi{1843.03} likely lies somewhere in between.

\subsection{Iron-enhanced USP planets}
\label{sec:FeUSPs}

\koi{1843.03} is one of a growing class of iron-enhanced, closely-orbiting planets discovered.

\begin{figure}
    \centering
    \includegraphics[width=\linewidth]{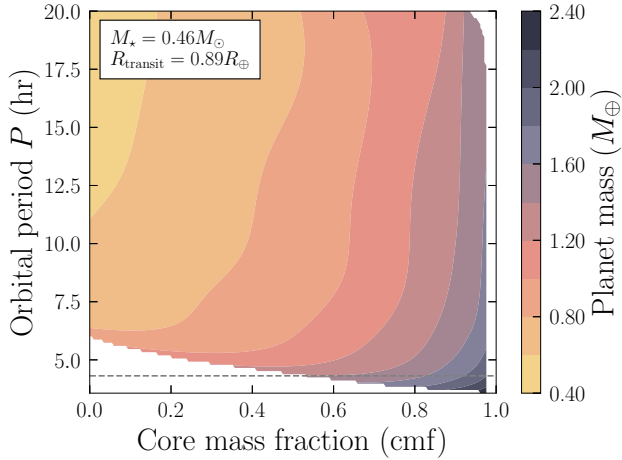}
    \caption{Contours of constant planet mass as a function of core mass fraction and orbital period for \ktwo{137b}. The orbital period of \ktwo{137b} is indicated by the dashed gray line. This figure is analogous to Figure~\ref{fig:contour-koi_1843}, but for the planet \ktwo{137b} instead of \koi{1843.03}.}
    \label{fig:contour-k2_137b}
\end{figure}

\begin{figure}
    \centering
    \includegraphics[width=\linewidth]{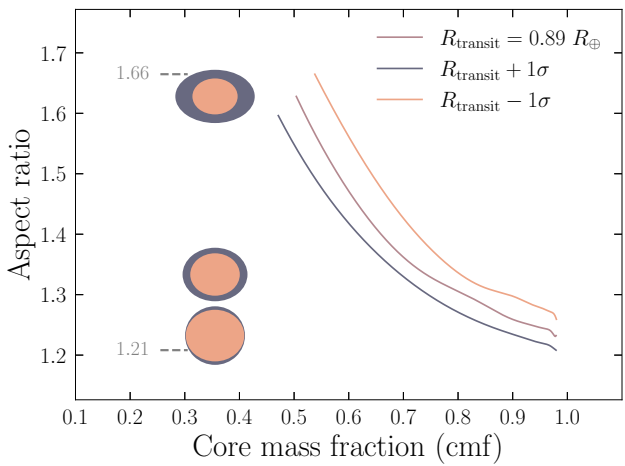}
    \caption{
    Aspect ratio constraints on \ktwo{137b}, as a function of core mass fraction at three different values for the transit radius: the measured value 0.89~\REarth and the $1\sigma$ limits (0.80~\REarth and 0.98~\REarth). This figure is analogous to Figure~\ref{fig:slices-aspect-koi_1843}, but for the planet \ktwo{137b} instead of \koi{1843.03}.}
    \label{fig:slices-aspect-k2_137b}
\end{figure}

\ktwo{137b} is remarkably similar to \koi{1843.03} but has a slightly longer orbital period (by 4 minutes), a larger transit radius ($0.89~\REarth$), and a more massive host star ($0.463~\MSun$) \citep{Smith+2018MNRAS}. Our 3D models show that \ktwo{137b} must be at least $42 \pm 5\%$ iron by mass to have avoided tidal disruption. Its mass lies somewhere between $1.01$ and $2.80~\MEarth$ (Figure~\ref{fig:contour-k2_137b}), which is consistent with the radial velocity upper limit of $3~\MJup$. We find that \ktwo{137b}'s aspect ratio is bounded between $1.21$ and $1.66$ (provided the true transit radius is within the 1$\sigma$ measurement, see Figure~\ref{fig:slices-aspect-k2_137b}). The constraint on \ktwo{137b}'s aspect ratio is not as extreme as the constraint on \koi{1843.03}'s. 

\begin{figure*}
    \centering
    \includegraphics[width=\linewidth]{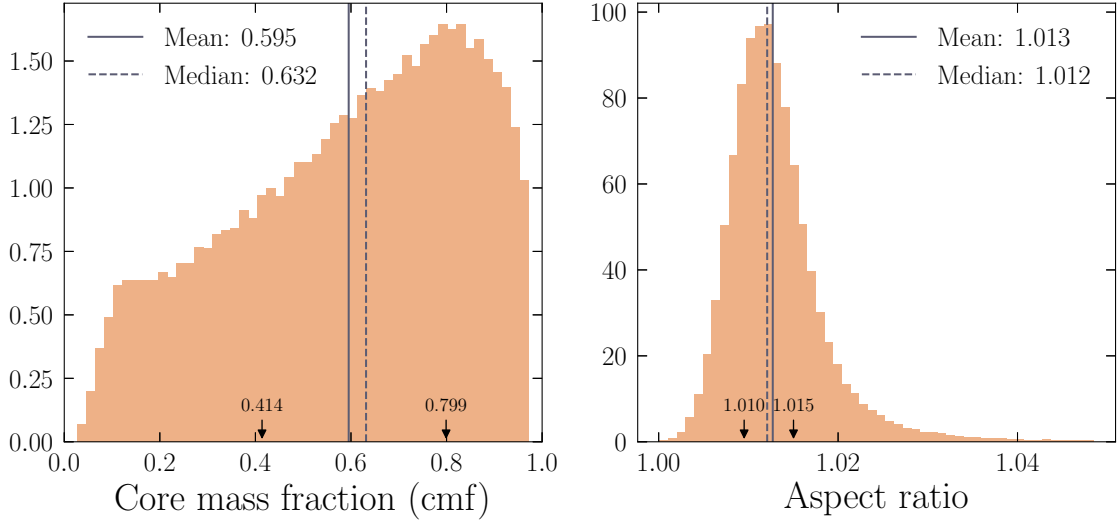}
    \caption{Histograms of core mass fraction and aspect ratio for \ktwo{229b}. The mean and median of each distribution are indicated by vertical lines, and black arrows show the 25\% and 75\% percentiles; we use these values instead of the standard deviation for errorbars in the text. We are not able to place a very tight constraint on the core mass fraction of this planet, but our models indicate that it is very unlikely to be significantly distorted. To compute these histograms, it is necessary to choose a mass and radius distribution; we select an uncorrelated bivariate Gaussian distribution based on the planet's measured mass and transit radius ($M_p = 2.59 \pm 0.43~\MEarth$ $R_p = 1.164^{+0.066}_{-0.048}~\REarth$, \citealt{Santerne+2018Nature}) for this purpose.}
    \label{fig:histograms-k2_229b}
\end{figure*}

\begin{figure*}
    \centering
    \includegraphics[width=\linewidth]{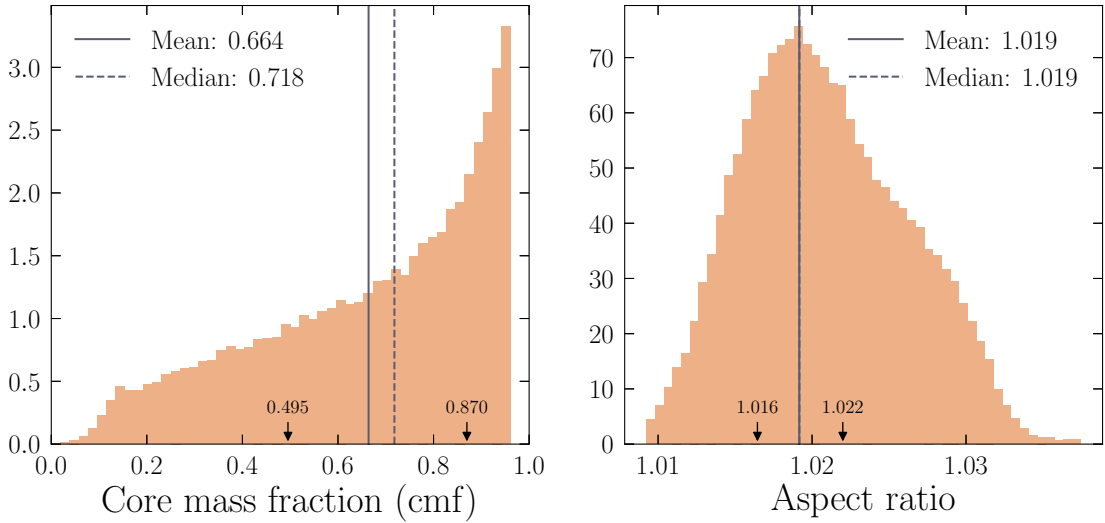}
    \caption{Histograms of core mass fraction and aspect ratio for \ktwo{106b} ($M_p = 8.36^{+0.96}_{-0.94}~\MEarth$ $R_p = 1.52 \pm 0.16~\REarth$, \citealt{Guenther+2017A&A}), analogous to Figure~\ref{fig:histograms-k2_229b}. We again find that, though the core mass fraction is not constrained very tightly, the aspect ratio is likely to be very close to unity. }
    \label{fig:histograms-k2_106b}
\end{figure*}

Two additional transiting exoplanets --- \ktwo{229b} ($M_p = 2.59~\MEarth$, $R_\mathrm{transit} = 1.164~\REarth$, $P_\mathrm{orb} = 14.0$~\hr, \citealt{Santerne+2018Nature}) and \ktwo{106b} ($M_p = 8.36~\MEarth$, $R_\mathrm{transit} = 1.52~\REarth$, $P_\mathrm{orb} = 13.7$~\hr, \citealt{Guenther+2017A&A}) --- have been inferred to have iron-rich compositions based on their radial velocity measured masses. Based on our models, we infer iron mass fractions of $0.605^{+0.204}_{-0.181}$ and $0.691^{+0.206}_{-0.169}$, and aspect ratios of $1.02^{+0.002}_{-0.003}$ and $1.01^{+0.003}_{-0.003}$, for \ktwo{229b} and \ktwo{106b} respectively. (Note that the errorbars reported here are not one standard deviation, as the distributions tend to be non-Gaussian, but rather the 25\% and 75\% percentiles; the full distributions are shown in Figures~\ref{fig:histograms-k2_229b} and \ref{fig:histograms-k2_106b}.)

Of the roughly 7 rocky ultra-short period exoplanets ($R_p \leq 1.7~\REarth$, $P_\mathrm{orb} \leq 1$~day) with masses and radii measured to within 20\% precision to date (CoRoT-7b, Kepler-10b, Kepler-78b, \ktwo{106b}, \ktwo{141b}, \ktwo{229b}, HD-3167b), 2 are iron-enhanced. Including the planets with density upper limits from the Roche limit (\koi{1843.03} and \ktwo{137b}), we find that just under half (4 out of 9) of the ultra-short period exoplanets with physically-meaningful constraints on their densities characterized to date are iron-enhanced.

\subsection{Analytic approximation to the Roche limit}

Our numerical Roche limits can be approximated by modifying the power-law parameters of the well-known expression for the classical Roche limit, re-expressed in terms of orbital period using Kepler's third law \citep{Rappaport+2013ApJL}; \revision{this is given by Equation~\ref{eqn:roche-mod}}. Including additional terms up to quadratic order that encapsulate a dependence on the degree of central concentration of the planet (i.e., the ratio of the planet's maximum/central density to its mean density), we obtain,
\begin{align}
    \log_{10}{\left(\frac{P_\mathrm{orb,min}}{1~\hr}\right)} &=& \log_{10}{C} + \alpha \log_{10}{x} + \beta \log_{10}{y} \nonumber \\
    & &{} + \gamma \left(\log_{10}{x}\right)^2 + \delta \left(\log_{10}{y}\right)^2 \nonumber \\
    & &{} + \eps \left(\log_{10}{x} \log_{10}{y}\right)
\end{align}
where $x \equiv \rho_p^{} / \left(1~\g~\cm[-3]\right)$, $y \equiv \rhomax / \rho_p^{}$. Using sequential least squares programming\footnote{SLSQP, implemented in SciPy, \citep{scipy}, using an algorithm originally developed by {Dieter Kraft}}, we find the following best fit parameters to the 2D surface defining the Roche limit: $C = 12.013$, $\alpha = -0.571$, $\beta = 0.047$, $\gamma = 0.108$, $\delta = 0.562$, and $\eps = -0.527$. 

\citet{Rappaport+2013ApJL} report an interpolation formula without the quadratic terms, where $C = 12.6$, $\alpha = -0.5$, and $\beta = -0.16$. If we set the quadratic terms to zero, our best-fitting revised Roche limit has $C = 11.86$, $\alpha = -0.52$, and $\beta = 0.086$.

While this approximation is a useful tool, interpolation within the model grid is our suggested approach for using the models.

%Other USP planet with periods < 7hours 
%KOI-4419.01 6.2hours, 1.22 Rearth
%\ktwo{141b} 6.7hrs but 1.51 Rearth so Roche limit places no constraints on rocky compositions 

\section{Discussion}
\label{sec:discussion}

\subsection{Formation scenarios}

How did \koi{1843.03}, \ktwo{137b}, \ktwo{229b} and \ktwo{106b} form and/or evolve to such short orbital periods and iron-rich compositions? 

\revision{Several theories for the origin of Mercury's high iron content involve impacts that remove the outer silicate layers of a differentiated planet. Scenarios involving a single giant impact \citep[e.g.,][]{Benz2008}, a hit-and-run collision \citep[e.g.,][]{AsphaugEt2014NatGeo}, or the cumulative effect of multiple collisions can all feasibly lead to a Mercury-like outcome, though a single-giant impact or hit-and-run impact require highly tuned collision geometries  to reproduce Mercury's mass and iron mass fraction \citep{ChauEt2018ApJ}. In the context of exoplanets, \citet{Marcus+2010ApJL} used smoothed particle hydrodynamics (SPH) simulations of giant impacts to infer that iron mass fractions of up to than 80\% can be achieved with a single impact for planets less than $2~\MEarth$. \citet{Marcus+2010ApJL} neglected to track the dynamical evolution of the impact ejecta, however,  and reaccretion of the ejected mantle is likely to dilute the iron-enhancement of a giant impact \citep{GladmanCoffey2009MPS}.} USP planets would be susceptible to high-velocity erosive collisions due to the extreme orbital velocities along their orbits (for example, about 320~\km~\s[-1] for \koi{1843.03} compared to about 48~\km~\s[-1] for Mercury, Table~\ref{tbl:compare}). \revision{Their proximity to their stars, with shorter orbital timescales and stronger stellar irradiation environments (Table~\ref{tbl:compare}), would also affect the reaccretion of ejected silicates. 
Further work is needed to evaluate the effect of collisions on the compositions of USP planets.} 

\begin{deluxetable}{lcc}
\tablehead{\colhead{Planet} & \colhead{Orbital velocity (\km~\s[-1])} & \colhead{Flux (\erg~\cm[-2]~\s[-1])}}
\tablecaption{Approximate values for the orbital velocities and fluxes at the orbital surface for the four planets we consider in this paper. We compare these values to the corresponding values for Mercury, the planet with the shortest orbital period in our own Solar System. \label{tbl:compare}}
\startdata
\koi{1843.03} & 320 & \expn{3}{9} \\
\ktwo{137b} & 270 & \expn{1}{9} \\
\ktwo{106b} & 270 & \expn{2}{10} \\
\ktwo{229b} & 240 & \expn{3}{9} \\
\tableline 
Mercury & 48 & \expn{9}{6}
\enddata
\end{deluxetable}

% A high-energy impact, one of the theories for the origin of Mercury's high iron content \citep{Benz2008}, may remove the outer silicate layers of a differentiated planet, provided the eroded material is not re-accreted by the planet \textbf{(as \citealt{GladmanCoffey2009MPS} found was likely in the case of  Mercury)} and is instead accreted by the star or blown away by radiation pressure. Smoothed particle hydrodynamics (SPH) simulations of giant impacts reveal iron mass fractions of more than 80\% can be achieved with a single impact \citep{Marcus+2010ApJL} for planets less than $2~\MEarth$. These ultra-short-period planets would be susceptible to high-velocity erosive collisions due to the extreme orbital velocities along their orbits (about 320~\km~\s[-1] for \koi{1843.03}, 266~\km~\s[-1] for \ktwo{137b}, 268~\km~\s[-1] for \ktwo{106b}, and 240~\km~\s[-1] for \ktwo{229b}, compared to about 48~\km~\s[-1] for Mercury).

Alternatively, these closely-orbiting iron-enhanced planets could have initially formed from iron-rich material. Both the condensation sequence (wherein iron condenses at a higher temperature than magnesium silicates) \citep{Lewis1972E&PSL} and photophoresis (which separates high-thermal-conductivity iron dust grains from lower-thermal-conductivity silicate grains) \citep{Wurm+2013ApJ} can lead to an enhancement of iron in the solid phase at the inner edge of the protoplanetary disk. These fractionation processes that operate primarily at the disk inner edge could imprint themselves as a statistical iron enhancement of the ultra-short-period planet population. 

A third possibility is that \koi{1843.03} and \ktwo{137b} are right at their Roche limits and have been gradually losing their outer silicate layers to Roche lobe overflow as their orbits tidally decay \citep{JiaSpruit2017MNRAS}. If \koi{1843.03} started with a chondritic or Earth-like iron-to-silicate ratio and mass of about $0.7~\MEarth$ (an intermediate value between our estimated limits), the planet's initial mass would have been about $1.4~\MEarth$. The orbital period precision achieved over the 4-year baseline of the \kepler\ mission is insufficient to resolve expected decay in \koi{1843.03}'s orbit. This scenario does not explain the compositions of the longer-orbital-period \ktwo{229b} and \ktwo{106b}, however, since they are outside their Roche limits.

\subsection{Thermal effects}

We have not modeled the interior temperature profiles of these planets. Indeed, we have adopted room temperature (300~\K) equations of state. The surface temperature of \koi{1843.03} and other USP planets can exceed 2000~\K, with temperature increasing further toward the center. Thermal expansion would cause the planet of specified mass and composition to have a larger volume and lower mean density compared to the models presented here. Temperature may also affect the pressure of the phase transition between enstatite and perovskite, which we have fixed to 23~GPa, following \citet{Sotin+2007Icarus}.  There are two common ways of incorporating temperature into the EOS \citep{Jackson1998}: One may either regard the typical EOS coefficients as being temperature-dependent, or one may add a ``thermal'' pressure at every point. 
The effect of temperature is more severe for lower mass bodies. Including thermal expansion will make the constraints on the iron mass fraction of \koi{1843.03} even more severe, strengthening our conclusions. 

\subsection{Effect of material strength}

%Intro Statement - state the facts.
\revision{Our models provide the first self-consistent constraints on the hydrostatic equilibrium shapes and Roche limits of  ultra-short period rocky planets. The effect of material strength in the planets' shapes is not taken into account in these calculations.}

%Our models assume the planets are in hydrostatic equilibrium, and neglect the influence of material strength in the planets' shapes.  

Looking to the Solar System bodies for inspiration, we see that once bodies are roughly 200~km (for icy materials) to 300~km (for rocky materials) in radius their self gravity is sufficient to overcome their material strength and they achieve a rounded shape. Iapetus (mean radius $734.5 \pm 2.8$~\km, \citealt{Roatsch2009}) is the largest Solar System body measured to have significant deviations from a hydrostatic equilibrium shape \citep{ThomasIcarus2010}. 
Since \koi{1843.03} is about 4000~km in radius, it is safely in the regime where self-gravity dominates the material forces and hydrostatic equilibrium determines its leading-order shape, \revision{satisfying the minimum mass criterion in the IAU definition of a planet}. 

\revision{To leading order, Earth-mass scale planets (such as \koi{1843.03}, \ktwo{137b}, \ktwo{229b}, and \ktwo{106b}) are in hydrostatic equilibrium, with rigidity representing a minor correction. In our models, the central pressure of \koi{1843.03} are on the order of $10^{11}$ -- $10^{12}$~Pa which is orders of magnitude larger than the shear strength of iron \citep{ClatterbuckEt2003AM} and peridotite \citep{Handy1999}.} %\citep[ideal shear strengths of  7.2~GPa and 7.8~GPa][]{ClatterbuckEt2003AM} and rock.
%The ideal shear strength (i.e., the stress at which the crystal lattice itself becomes unstable) has been computed to be 7.2~GPa and 7.8~GPa for iron \citep{ClatterbuckEt2003AM}. The ideal strength sets a firm upper limit on the strength of a crystalline material; defects typically cause materials to deform at stresses well below the ideal strength.} 

\revision{The high instellations of USP planets can lead to molten surfaces \citep[e.g.,][]{Leger+2011Icarus, Kite+2016ApJ}, which further limit deviations from hydrostatic equilibrium shapes.} In the extreme of no heat redistribution, the substellar point of \koi{1843.03} could exceed 2000~\K, computed from $T_{ss} = T_\star \left(1 - \alpha\right)^{1/4} \sqrt{R_\star / a} \approx 2500$~\K\, \revision{assuming a basalt-like planet surface albedo} $\alpha = 0.1$ \citep[e.g.,][]{Kite+2016ApJ}. This temperature is sufficiently hot \revision{to melt metallic iron (melting point 1811~K) and is hotter than the liquidus of peridotite \citep{Takahashi1986JGR}, the dominant rock in Earth's upper mantle. 
%to melt iron and enstatite (with melting temperature 1831~\K\ at $P = 0$, \citealt{Boyd+1964JGR,Belonoshko+2005PRL}).
Thus, the planet's surface would have too little strength to sustain topography that could significantly influence the transit depth.}

%Add paragraph a la Josh Winn re survival of planets inside the Roche limit. 
\revision{In using the Roche limit to constrain the bulk compositions of \koi{1843.03} and \ktwo{137b}, we have followed \citet{Rappaport+2013ApJL} and \citet{JiaSpruit2017MNRAS} and neglected the effect of material strength.
It is unclear whether material strength or friction would help the planet to survive intact inside its Roche limit for gigayear timescales \citep[e.g.,][]{Davidsson1999Icar, HolsappleEt2006Icar}. 
As highlighted by \citet{WinnEt2018NewAR}, further work is needed to model the destruction of USP planets that exceed their Roche limits.}

\subsection{Planet mass loss}

The surface of \koi{1843.03} could be actively sublimating. \citet{Kite+2016ApJ} models the exchange between atmospheric silicate, surface magma pools, and interior material for a hot, rocky exoplanet. If a rock vapor atmosphere is contributing to the transit depth in the \kepler\ bandpass, that only makes our constraints on the the iron fraction in \koi{1843.03} even more severe. 

Though they do not themselves show evidence of evaporation in \kepler\ photometry, \koi{1843.03}, \ktwo{137b}, \ktwo{229b} and \ktwo{106b} could be more massive cousins to the catastrophically evaporating rocky planet discovered orbiting \kic{12557548}. \kic{12557548} shows asymmetric and variable transit shapes that have been interpreted as evidence of a dusty outflow of vaporized material driven by a thermal wind \citep{Rappaport+2012ApJ}. Even for \koi{1843.03}, the smallest among these close-orbiting iron-enhanced planets, with a mass in excess of $0.3~\MEarth$, the escape velocity from the surface is too high to drive a substantial hydrodynamic wind of sublimated silicates \citep{Perez-Becker&Chiang2013}. Detailed models of radiative hydrodynamic winds from evaporating rocky USP planets \citep{Perez-Becker&Chiang2013} show that a $0.1~\MEarth$ rocky planet could survive at a surface temperature of $\sim 2200$~K with negligible mass loss for tens of gigayears.

%Though it does not itself show evidence of evaporation, \koi{1843.03} could be a more massive cousin to the catastrophically evaporating rocky planet discovered orbiting \kic{12557548} \citep{Rappaport+2012ApJ}. \kic{12557548} --- as well as \koi{2700b} \citep{Rappaport+2014ApJ} and \ktwo{22b} \citep{SanchisOjeda+2015ApJ} --- show asymmetric and variable transit shapes that have been interpreted as evidence of a dusty outflow of vaporized material driven by a thermal wind. \koi{1843.03}, on the other hand, does not show any evidence of a comet-like tail in the \kepler\ photometry. This is to be expected because more massive rocky planets are less susceptible to thermally-driven outflows due to their higher escape velocities. The observed comet-like evaporating planets are estimated to have masses $< 3~\MJup$, but modelling by \citet{Perez-Becker&Chiang2013} constrains the masses to be roughly $< 1~\MEarth$. We estimate a minimum mass of $0.33~\MEarth$ for \koi{1843.03}. Models of radiative hydrodynamic winds from evaporating rocky USP planets have shown that a $0.1~\MEarth$ rocky planet could survive at a surface temperature of 2200~K with negligible mass loss for tens of gigayears \citep{Perez-Becker&Chiang2013}.

\subsection{Potential for follow-up observations}

Due to the red colors of the early M host-star, the near infrared (NIR) and infrared (IR) offer the best opportunities for further observational characterization of \koi{1843.03}. Given our constraints on its mass and bulk composition, the possible range for \koi{1843}'s radial velocity semi-amplitude ($K_1 = 0.60$ -- $1.98$~m~s${}^{-1}$) spans the current state of the art precision of $1$~\m~\s[-1]. The host star is too faint for precision radial velocity follow-up in the visible with any existing telescope or instrument but may be a feasible candidate for radial velocity follow-up in the NIR.  Photometric follow-up in the infrared, for instance with the {\it Spitzer} Space Telescope or the James Webb Space Telescope could confirm the planetary nature of \koi{1843.03} \citep{Desert+2015ApJ}. 

Photometric follow-up could also provide a longer time baseline to reveal evidence of tidal evolution of the orbit. In principle, given sufficient time sampling, the detailed shape of the transit lightcurve (as the projected cross-section of the planet changes viewing angle during transit) may further constrain \koi{1843.03}'s aspect ratio and bulk composition. Such an effect has been studied for distorted giant planets \citep{Leconte+A&A201}.

To predict the IR transit signal-to-noise ratio (SNR), we use the \texttt{isochrones} software \citep{Morton2015ASCL} modified for \textit{Spitzer} bandpasses using data from \citet{Hora+2008PASP} and \citet{Indebetouw+2005ApJ} to compute predicted apparent magnitudes for \koi{1843} in each of the four {\it Spitzer} bandpasses. In the 3.6~\um, 4.5~\um, 5.8~\um, and 8.0~\um bands, respectively, we predict magnitudes of $11.0 \pm 0.05$, $11.0 \pm 0.08$, $10.9 \pm 0.08$, and $10.9 \pm 0.07$. Scaling from the SNRs obtained by \citet{Desert+2015ApJ} on stars with similar \textit{Spitzer} magnitudes as \koi{1843}, a signal-to-noise ratio for a single \koi{1843.03} transit could range from about $0.36$ to $0.93$. Observing multiple transits could improve the SNR.

\koi{1843.03} may not hold the records for the shortest orbital period and most distorted known exoplanet for long. The Transiting Exoplanet Survey Satellite (TESS, launched in 2018) should find several ultra-short-period transiting planets as it surveys the brightest stars over the entire sky; recent simulations by \citet{Barclay+2018ApJS} predict detection for $52$ planets with orbital periods of $P_\mathrm{orb} < 1$~day. Our models, which provide the first self-consistent constraints on the Roche limits of Earth-mass-scale rocky planets (Figure~\ref{fig:minperiod-fe}), will enable composition constraints on these future ultra-short period planet discoveries. 

\software{george \citep{george}, GNU Scientific Library \citep{gsl}, isochrones \citep{Morton2015ASCL}, scipy \citep{scipy}}

\acknowledgements 

E.M.P. and L.A.R. thank Dr. Benjamin T. Montet for his analysis of the \koi{1843.03} \kepler\ light curve and and constraint on the rate of change in the planet's orbital period. We also thank Drs. Andrew Vanderburg, Saul Rappaport, Josh Winn, and Darin Ragozzine for helpful discussion, and reviewer Dr. Erik Asphaug for valuable suggestions.

This material is based upon work supported by a National Science Foundation Graduate Research Fellowship under Grant Nos. DGE1144152 and DGE1745303. L.A.R. acknowledges NSF grant AST-1615315. The computations in this work were carried out with resources provided by the University of Chicago Research Computing Center.

\bibliographystyle{aasjournal}
\bibliography{biblio}

\end{document}